\documentclass[twocolumn]{aastex62}

\usepackage{graphicx}
\newcommand\teff{$T_{\mathrm{eff}}$}
\newcommand\teffq{$T_{\mathrm{eff}}^4$}
\newcommand{\gf}{$g_{F}$} 
\newcommand\logg{$\log\,g$}
\newcommand\loggf{$\log g_{F}$}
\newcommand\loggfeq{\log g_{F}}

\newcommand\Mbol{$M_{\mathrm{bol}}$}
\newcommand\Msun{$M_{\odot}\;$}
 
\submitjournal{ApJ}

\shorttitle{Modified gravity and BSG FGLR}
\shortauthors{Sextl et al.}

\begin{document}

\title{Modified Gravity and the Flux-weighted Gravity-Luminosity Relationship of Blue Supergiant Stars}

\correspondingauthor{Rolf-Peter Kudritzki}
\email{kud@ifa.hawaii.edu}

\author{Eva Sextl}
\affiliation{Universit\"ats-Sternwarte, Fakult\"at f\"ur Physik, Ludwig-Maximilians Universit\"at M\"unchen, Scheinerstr. 1, 81679 M\"unchen, Germany}
\author{Rolf-Peter Kudritzki}
\affiliation{Universit\"ats-Sternwarte, Fakult\"at f\"ur Physik, Ludwig-Maximilians Universit\"at M\"unchen, Scheinerstr. 1, 81679 M\"unchen, Germany}
\affiliation{Institute for Astronomy, University of Hawaii at Manoa, 2680 Woodlawn Drive, Honolulu, HI 96822, USA}
\author{Jochen Weller}
\affiliation{Universit\"ats-Sternwarte, Fakult\"at f\"ur Physik, Ludwig-Maximilians Universit\"at M\"unchen, Scheinerstr. 1, 81679 M\"unchen, Germany}
\affiliation{Max Planck Institute for Extraterrestrial Physics, Giessenbachstrasse, 85748 Garching, Germany}
\author{Miguel A.\ Urbaneja}
\affiliation{Institut f\"ur Astro- und Teilchenphysik, Universit\"at Innsbruck, Technikerstr. 25/8, 6020 Innsbruck, Austria}
\author{Achim Weiss}
\affiliation{Max Planck Institute for Astrophysics, Karl-Schwarzschild-Str. 1, 85748 Garching, Germany}
\

\begin{abstract}
  We calculate models of stellar evolution for very massive stars and include the effects of modified gravity to investigate the influence on the physical properties of blue supergiant stars and their use as extragalactic distance indicators. With shielding and fifth force parameters in a similar range as in previous studies of Cepheid and tip of the red giant branch (TRGB) stars we find clear effects on stellar luminosity and flux-weighted gravity. The relationship between  flux weighted gravity, \gf $\equiv$ g/\teffq, and bolometric magnitude \Mbol~(FGLR), which has been used successfully for accurate distance determinations, is systematically affected. While the stellar evolution FGLRs show a systematic offset from the observed relation, we can use the differential shifts between models with Newtonian and modified gravity to estimate the influence on FGLR distance determinations. Modified gravity leads to a distance increase of 0.05 to 0.15 magnitudes in distance modulus. These change are comparable to the ones found for Cepheid stars. We compare observed FGLR and TRGB distances of nine galaxies to constrain the free parameters of modified gravity. Not accounting for systematic differences between TRGB and FGLR distances shielding parameters of 5$\times$10$^{-7}$ and 10$^{-6}$ and fifth force parameters of 1/3 and 1 can be ruled out with about 90\%  confidence. Allowing for potential systematic offsets between TRGB and FGLR distances no determination is possible for a shielding parameter of 10$^{-6}$. For 5$\times$10$^{-7}$ a fifth force parameter of 1 can be ruled out to 92\% but 1/3 is unlikely only to 60\%.
  
\end{abstract}


\section{Introduction}

The discovery of dark matter and of the accelerated expansion of the universe has triggered a substantial amount of approaches to explain these revolutionary astronomical discoveries by a modification of general relativity. These approaches introduce a new scalar degree of freedom in scalar-tensor theories coupling to ordinary matter and lead to a fifth force which changes gravitational attraction. Most promising are concepts which include a screening mechanism where the new force is suppressed in deep potential wells or regions of high density but influences gravity outside these regions. A prominent example of this class of models are the ones discussed by \citet{2004PhRvD..69d4026K} and \citet{2004PhRvD..70l3518B}, where a non-linear screening of a scalar field, a so called chameleon field, can suppress modifications of gravity on galactic scales. For an introduction and overview we refer to \citet{Chang2011, Davis2012, Jain2013, Desmond2019, Sakstein2020}. In the case of massive galaxies or of dwarf galaxies in the neighborhood of massive galaxies the potential wells will shield the stars against the fifth force (see \citealt{Cabre2012} for a detailed description), but for isolated smaller galaxies modified gravity may affect the internal structure of stars through a modification of the equation of hydrostatic equilibrium resulting in changes of observational stellar properties such as luminosity, temperature, radius and pulsation periods. Most promising are evolved stars, giants or supergiants, because their envelopes may be unscreened against the fifth force due to their large radii and the resulting lower gravitational potential.

\citet{Jain2013} systematically investigated two types of evolved stars, low mass stars at the tip of the red giant branch (TRGB) and massive Cepheid stars of 5 to 10 \Msun~in the stellar instability strip. They found that both the luminosity of TRGB stars and the period-luminosity relationship of Cepheids are affected. As a result, extragalactic distances obtained from these objects are altered, most interestingly in opposite direction, as TRGB distances decrease due to a decreased luminosity (see also \citealt{Desmond2020a}), whereas Cepheid distances increase because of the decrease of pulsation period. Comparing with observed galaxy distances \citet{Jain2013} obtained constraints on the potential scalar background field.    

Blue supergiant stars provide an important alternative to Cepheids as distance indicators through their relationship between flux-weighted gravity and luminosity \citep{Kudritzki2003, Kudritzki2008, Urbaneja2017}. They have higher masses than Cepheids and are significantly more luminous with substantially larger radii. The gravitational potential of their envelopes is comparable to the one of Cepheids and, consequently, effects of modified gravity may be equally important. We have therefore carried out stellar evolution calculations for massive stars in the range from 12 to 60 \Msun~to investigate how modified gravity influences the observable stellar properties and whether it introduces detectable systematic changes to the flux-weighted gravity - luminosity relationship (FGLR).

 \begin{figure}[ht!]
  \begin{center}
    \includegraphics[scale=0.60]{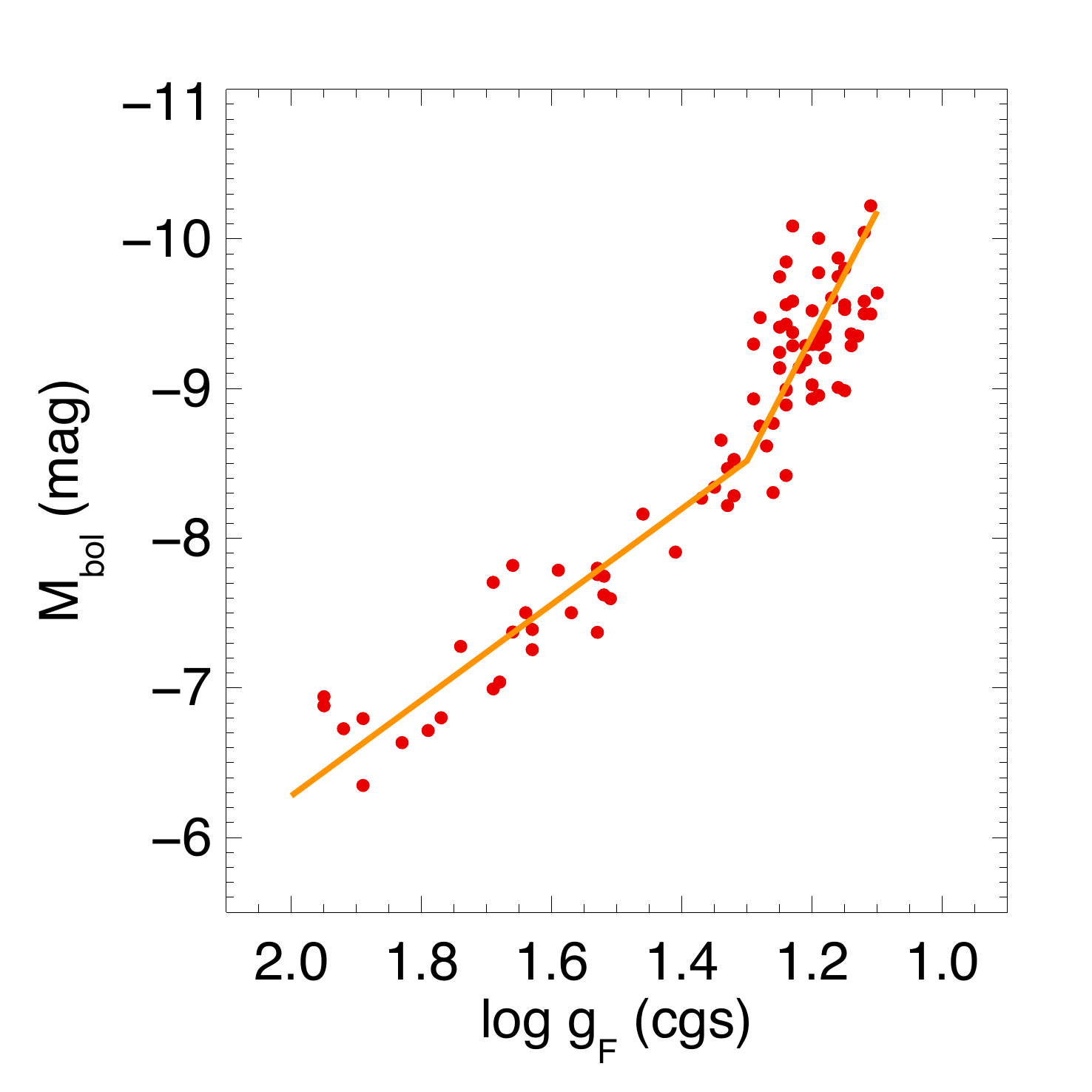}
  \end{center}
  \caption{
The observed FGLR of blue supergiant stars in the LMC (see text). 
} \label{figure_1}
\end{figure}

\section{Flux-weighted Gravity- Luminosity Relationship (FGLR)}

Blue supergiant stars (BSG) are massive stars in the temperature range 7900K $\leq$ \teff $\leq$ 25000K (see \citealt{Urbaneja2017}). Their tight observational relationship between absolute stellar magnitude \Mbol~and flux weighted gravity \loggf, \loggf~= \logg~- 4log(\teff/10$^4$K), has been discovered by \citet{Kudritzki2003}. It is a simple consequence of the well known power law relationship between stellar luminosity and mass and the fact the massive stars evolve from their hydrogen burning main sequence to the red supergiant phase at almost constant luminosity. With absolute visual magnitudes up to -10 mag BSG are beacons in the universe and much brighter than other stellar distance indicators such as Cepheids or TRGB stars.  This allows for spectroscopic studies out to distances of 10 Mpc with present day telescopes and a factor four further out with the next generation of 30m to 40m  ground based telescopes. In consequence, the FGLR has a great potential for extragalactic distance determinations. \citet{Kudritzki2008} studying a large sample of BSG in the sculptor galaxy NGC 300 have provided a first calibration of the FGLR and subsequently distances to eight galaxies have been determined: WLM -- \citet{Urbaneja2008}; M33 -- \citet{U2009}; M81 -- \citet{Kudritzki2012}; NGC3109 -- \citet{Hosek2014}; NGC3621 -- \citet{Kudritzki2014}; M83 -- \citet{Bresolin2016}; NGC55 -- \citet{Kudritzki2016}; IC1613 -- \citet{Berger2018}. We note that these galaxies cover a wide range of stellar metallicities from about 1/10 solar to twice solar but no significant metallicity dependence of the FGLR has been found. These observational findings are in agreement with stellar evolution calculations which also show that the effects of metallicity on the FGLR are small \citep{Meynet2015}.

The most recent calibration of the FGLR is given by the work of \citet{Urbaneja2017}. They carried out a detailed quantitative spectroscopic NLTE analysis of 90 BSG in the LMC and determined stellar effective temperatures, gravities and element abundances. Figure \ref{figure_1} shows the FGLR resulting from their work. The 2-component regression fit to the data provides the new calibration, which is given by

\begin{equation}
M_{\mathrm{bol}} = a(\loggfeq - 1.5) + b
\end{equation}
if $\loggfeq \geq \loggfeq^{\mathrm{break}}$, and
\begin{equation}
M_{\mathrm{bol}} = a_{\mathrm{low}}(\loggfeq - \loggfeq^{\mathrm{break}}) + b_{\mathrm{break}}
\end{equation}
if $\loggfeq \leq \loggfeq^{\mathrm{break}}$, with
\begin{equation}
b_{\mathrm{break}} = a(\loggfeq^{\mathrm{break}} - 1.5) + b,
\end{equation}
where $\loggfeq^{\mathrm{break}}$ = 1.30 dex, $a$ = 3.20 $\pm$ 0.08, $b$ = --7.878 $\pm$ 0.02 mag, and $a_{\mathrm{low}}$ = 8.34 $\pm$ 0.25.

We note that our Figure \ref{figure_1} and the value of b in eqn. (1) are slightly different from the original results obtained by \citet{Urbaneja2017}, since we now use the 1 percent precision distance to the LMC determined by \citet{Pietrzynski2019} from the light curve and radial velocity analysis of 20 late type eclipsing binaries and an improved stellar surface brightness-color relationship. 

\section{Modified Gravity and Massive Star Evolution}

The effects of modified gravity on stellar structure depend on two free parameters, the self-screening parameter $\chi_c$ and the fifth force parameter $\alpha_c$ (see \citealt{Jain2013}). $\chi_c$ describes how efficient a star is screening itself against the fifth force. It is used to determine the screening radius r$_s$ inside the star through the condition

\begin{equation}
  \chi_c = {4\pi \over c^2}  G_0 \int_{r_s}^R r \rho(r) dr,
\end{equation}

where r is the radial coordinate inside the star, R the stellar radius, G$_0$ the Newtonian gravitational constant and c the speed of light. Inside the screening radius the fifth force is screened and only Newtonian gravity with G$_0$ is acting. $\rho(r)$ is the density profile of the star. Exterior to r$_s$ the fifth force contributes and leads to a radius dependent gravitation via

\begin{equation}
  G(r) = G_0 \left[1 + \alpha_c \left(1 - {M(r_s) \over M(r)}\right)\right].
\end{equation}  

$\alpha_c$ sets the maximum contribution of the fifth force. M(r) and M(r$_s$) correspond to the stellar mass enclosed inside the radius r and the screening radius r$_s$, respectively.

For the physics of stellar structure and evolution the implementation of modified gravity through eqns. (4) and (5) is straightforward through a replacement of G$_0$ by G(r) in the equation of hydrostatic equilibrium as demonstrated in the work by \citet{Chang2011, Davis2012, Jain2013}. For main sequence stars, analytical estimates of the main effects can easily be obtained. For instance, using the well know mass-luminosity relation (see eq. 6 below) it is straightforward to show that stellar luminosity increases as a consequence of the rise from G$_0$ to G(r). For advanced stages of stellar evolution such as Cepheids and red giants stars numerical models are needed in conjunction with analytical considerations.

The case of massive BSG is more complex, because their evolution is complicated by the effects of strong stellar winds and rapid rotation. Thus, to investigate the influence of modified gravity  requires the detailed use of numerical models. For our study we have used the \texttt{MESA} stellar evolution code version 12115 \citep{Paxton2011, Paxton2013, Paxton2015, Paxton2018, Paxton2019}. One of the many advantages of \texttt{MESA} is that it provides a simple way to override most of physical routines without modifying more than one file. Since \texttt{MESA} already has a variable that holds the value of the gravitational constant for each radial cell in the star, the implication of modified gravity for a given pair of values $\chi_c$ and $\alpha_c$ is simple. All it requires is a numerical determination of the screening radius r$_s$ and then the values of G$_0$ are replaced by G(r) outside the screening radius in all equations, which contain G. This approach has already been used by \citet{Chang2011, Davis2012, Jain2013}.

For our numerical calculations of BSG evolution using \texttt{MESA} we adopt Z = 0.0067 for the metallicity mass fraction corresponding to the average metallicity of the 90 BSG in the LMC determined by \citet{Urbaneja2017}. For the effects of mass-loss we apply the \texttt{MESA} module, which uses the results obtained by \citet{Vink1999, Vink2000, Vink2001} based on the theory of radiation driven winds (see \citealt{Kudritzki2000}). For internal stellar layers with convection a mixing length parameter $\alpha_{\rm MLT}$ = 1.6 and step function overshooting with an overshooting parameter $\alpha_{\rm OV}$ = 0.1 are used. Our models include stellar rotation with initial rotational velocities on the zero-age-main-sequence as in  \citet{Ekstrom2012} and they also account for rotationally enhanced mass-loss. Eddington-Sweet circulation (with Eddington Sweet factor 0.5) instability, Goldreich-Schubert-Fricke instability, Solberg-Hoiland instability and secular shear instability \citep{Heger2000} induced by rotation and resulting in mixing are also included.

For our investigation we calculate a grid of evolutionary models with initial masses of 12, 13, 15, 18, 20, 23, 25, 29, 32, 37, 40, 43, 45, 50, 60 \Msun, respectively. To check our adaption of the \texttt{MESA} code we also compare in detail with the comprehensive sets of the Newtonian state-of-the-art stellar evolution models by \citet{Ekstrom2012} and \citet{Georgy2013}. For this comparison we adjust our metallity to their values (Milky Way and Small Magellanic Cloud metallicities, respectively). We find good agreement over the whole mass range.

For the selection of the modified gravity parameters $\alpha_c$ and $\chi_c$ we use the work by \citet{Jain2013} as guideline. \citet{Jain2013} discussed self-shielding parameters $\chi_c$ in a range from 10$^{-7}$ to 10$^{-6}$ and fifth force parameters $\alpha_c$ in a range from 0.2 to 1 with special emphasis on models with $\alpha_c$ = 1/3 and 1. Comparing observed galaxy distances determined from Cepheid and TRGB stars  they concluded that for $\alpha_c$ = 1/3 values of  $\chi_c \geqq$ 5$\times$10$^{-7}$ can be ruled out with 95\% confidence. For $\alpha_c$ = 1 values of $\chi_c \geqq$ 1$\times$10$^{-7}$ can be ruled out with similar evidence. Following this discussion we select $\chi_c$ = 5$\times$10$^{-7}$ and 10$^{-6}$ and $\alpha_c$ = 1/3 and 1, respectively, and compare with models which do not include modified gravity. We note that $\alpha_c=1/3$ corresponds to models with non-linear terms in gravity sector of the Lagrangian, i.e. so called $f(R)$ theories, as for example explored as inflationary scenario by \citet{1980PhLB...91...99S} or in the context of dark energy by \citet{2007PhRvD..76f4004H} or additional spatial dimensions \citep{2000PhLB..485..208D}, albeit the screening mechanism in the latter is the Vainshtein mechanism \citep{Vainshtein1972}. We should additionally note that a recent analysis by \citet{Desmond2020b}, who investigated the constraints on modified gravity from a statistical study of  galaxy morphology, seems to rule out Hu-Sawacki f(R) gravity.

\begin{figure*}[ht!]
  \includegraphics[scale=1.10]{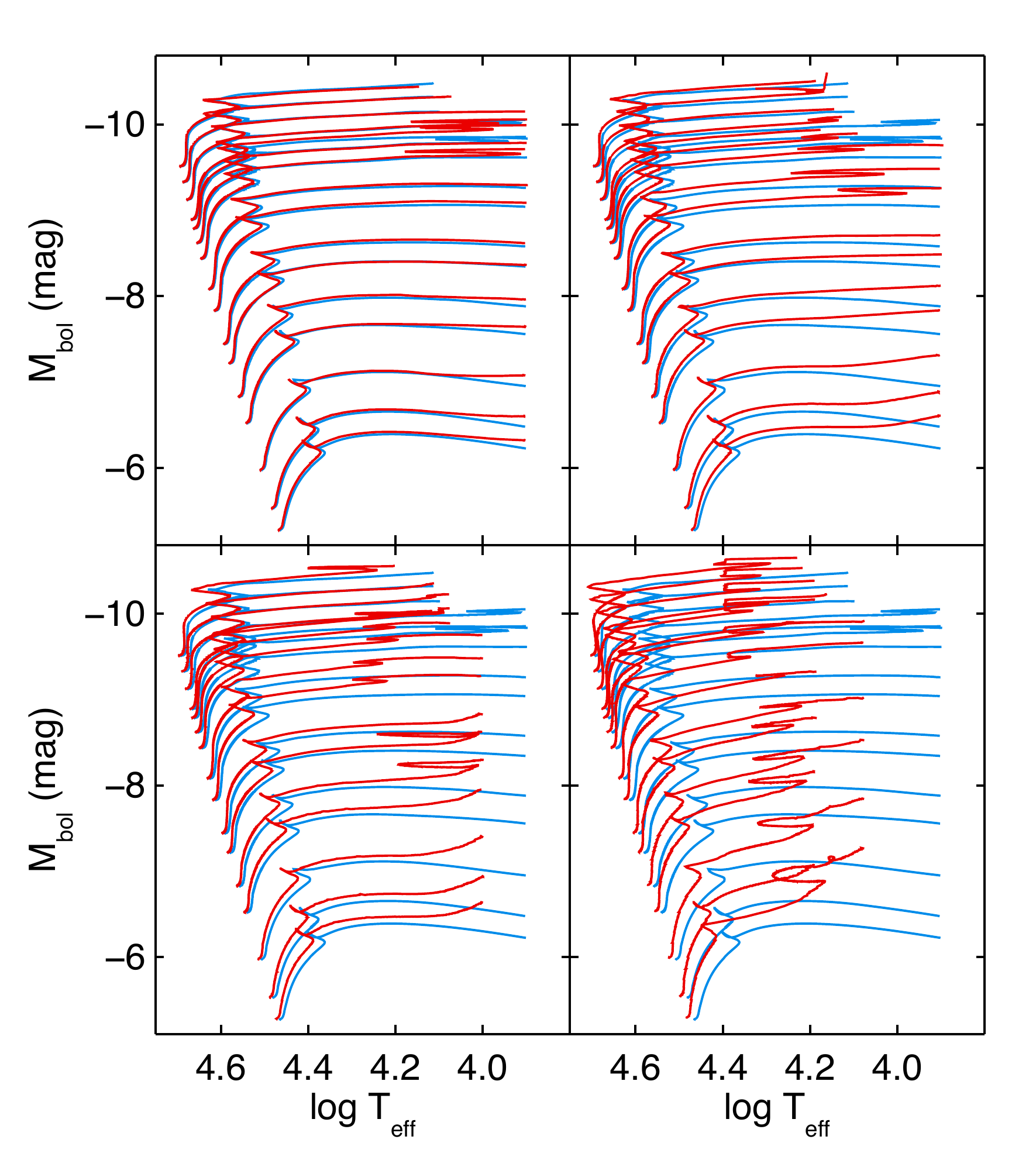}
\caption{
Stellar evolution in the HRD. Evolutionary tracks starting at the ZAMS and ending at the end of the BSG phase with \teff~= 7900K are shown for initial masses from 12 to 60 \Msun, respectively (see text for the masses of each track). Tracks in red include the effects of modified gravity, while tracks in blue are calculated with solely Newtonian gravity. Top left: modified gravity with $\alpha_c$ = 1/3 and $\chi_c$ = 5$\times$10$^{-7}$; top right: $\alpha_c$ = 1/3 and $\chi_c$ = 10$^{-6}$; bottom left: $\alpha_c$ = 1 and $\chi_c$ = 5$\times$10$^{-7}$; bottom right: $\alpha_c$ = 1 and $\chi_c$ = 10$^{-6}$.
}
\label{figure_2}
\end{figure*}

Figure \ref{figure_2}  shows the  classical Hertzsprung-Russell diagram (HRD) of the evolutionary tracks for both cases, Newtonian and modified gravity. As we can see, modified gravity increases the luminosity of massive stars. The effect increases with screening parameter $\chi_c$ and fifth force parameter $\alpha_c$. This effect is a simple consequence of the change from G$_0$ to G(r) as described by equation (5).

As is well known (see, for instance, \citealt{Kippenhahn2012} or \citealt{Davis2012}), a simple estimate of the relationship between stellar luminosity and stellar mass for stars on the main sequence yields

\begin{equation}
L \propto G^4M^3.
\end{equation}  

We note that in our range of stellar masses the exponent of M is slightly smaller, 2.5 instead of 3, because of the influence of radiation pressure which increases with increasing stellar mass. We also note from Figure \ref{figure_2} that BSG follow a similar relationship except that the luminosities are about a factor of ten larger than on the main sequence. Therefore, if the screening radius is located deep inside the massive stars, they should experience the increase from G$_0$ to G(r) and become more luminous. In the following we discuss this more quantitatively. We introduce an analytical fit of the density stratification inside massive stars during their evolution and and use this to derive an approximation for the location of the screening radius as a function of stellar parameters.

The density distribution of a massive star on the main sequence is reasonably well approximated by adopting constant density inside the stellar core r$_c$ and an exponential decline with scale height H outside the stellar core: 

\begin {equation}
  \rho(r) = \rho_c^{ms},~~r \leq r_c, ~~r_c = f_{\delta}^{ms}H, ~~H=\delta_{ms} R,
\end{equation}
\begin{equation}
 \rho(r) = \rho_c^{ms}e^{{r_c-r} \over H}~~~~~r_c \leq r \leq R,
\end{equation}

and

\begin{equation}
  \rho_c^{ms} = {M \over {4\pi (\delta_{ms} H)^3}}\times {1 \over \xi_{ms}}
\end{equation}

with

\begin{equation}
  \xi_{ms} = {(f_{\delta}^{ms})^3 \over 3}+2[(1+{r_c \over H} + {r_c^2 \over 2H^2}) - e^{r_c - R \over H}(1+{1 \over \delta_{ms}} + {1 \over 2\delta_{ms}^2})].
\end{equation}    

\begin{figure*}[ht!]
  \includegraphics[scale=0.6]{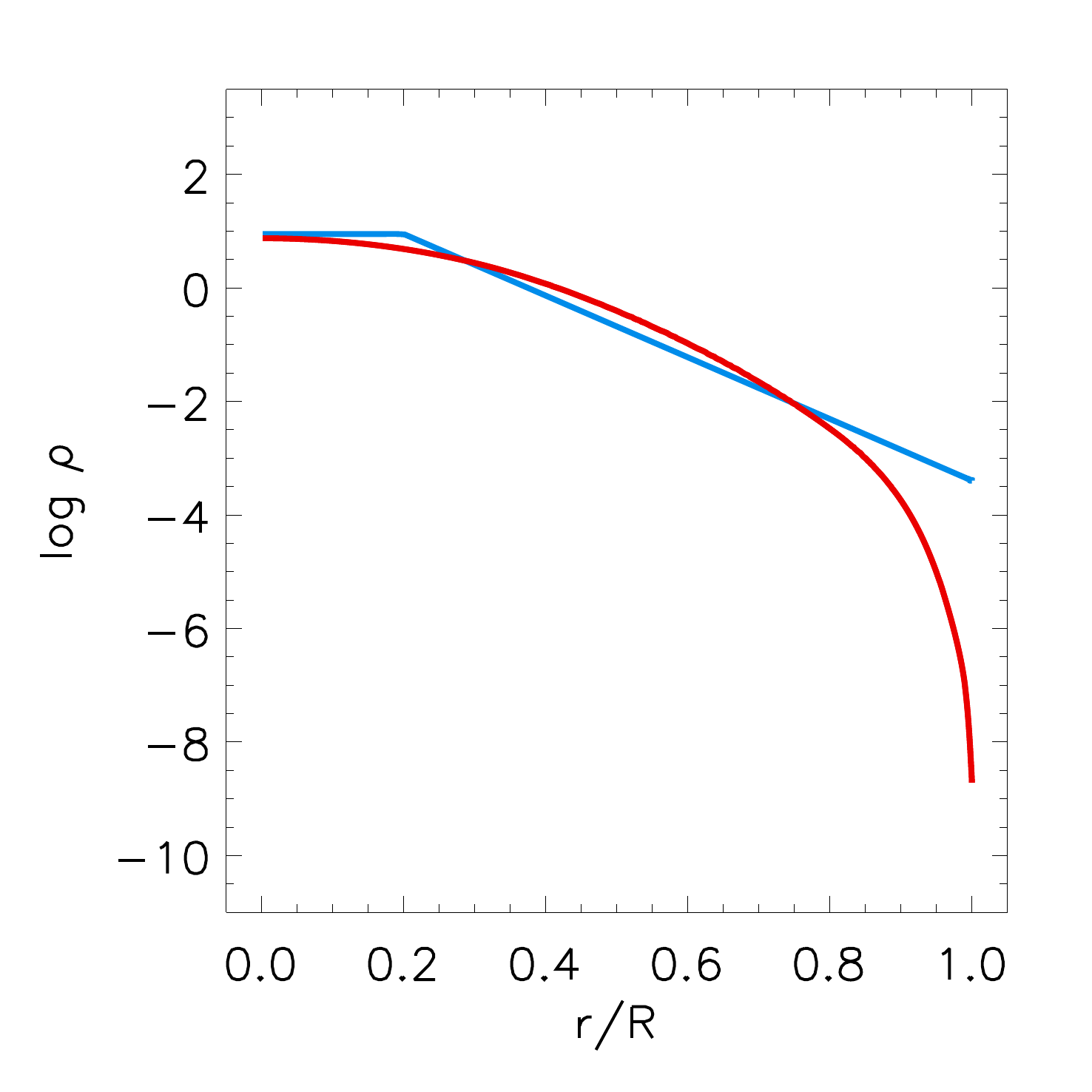}
  \includegraphics[scale=0.6]{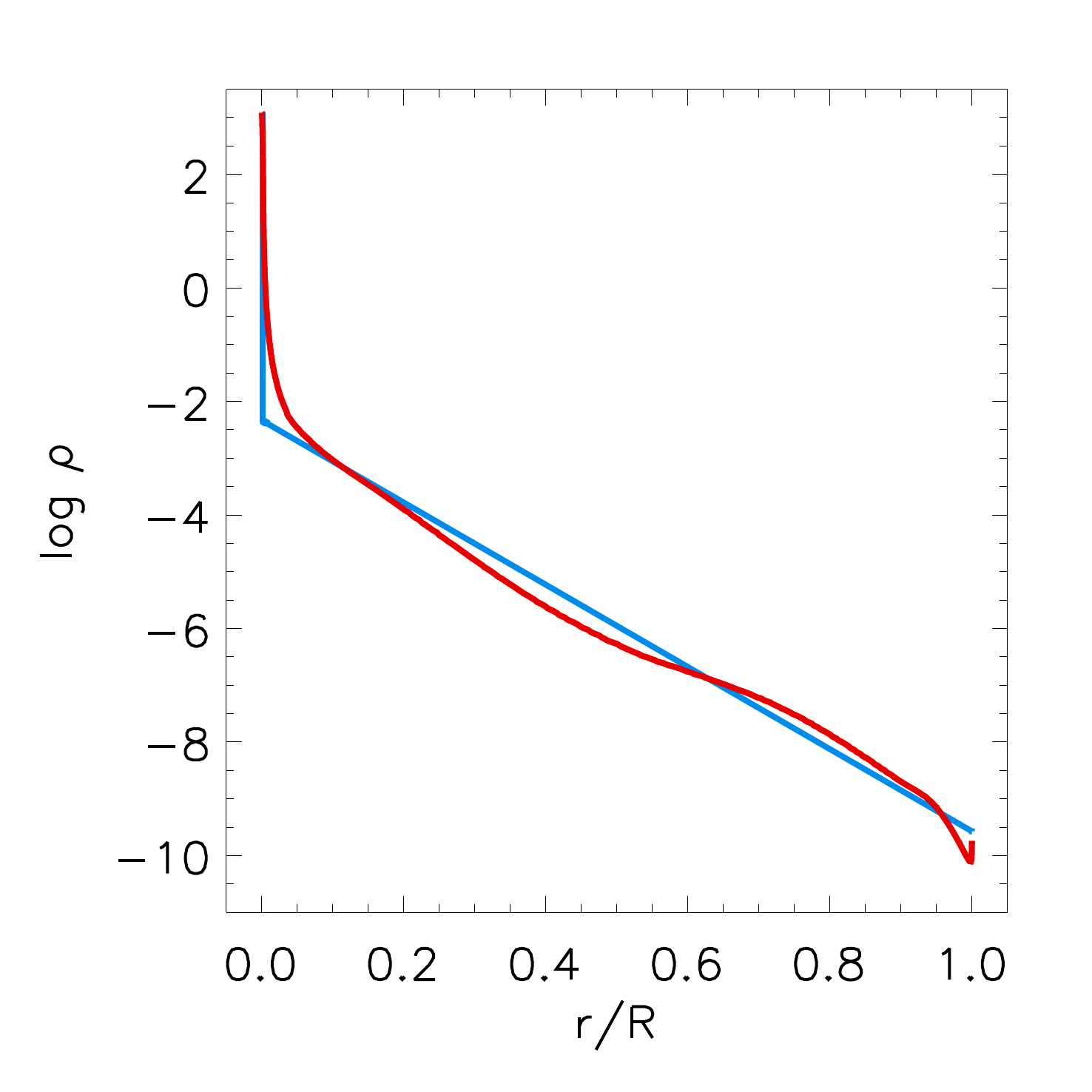}
  \caption{
Density distributions of MESA stellar evolution model of a star with 15 \Msun~(red). Left: ZAMS; right: BSG phase at \teff = 7900K. The analytic fits described in the text are shown as blue curves.  
}
\label{figure_3}
\end{figure*}

\begin{figure*}[ht!]
  \includegraphics[scale=0.60]{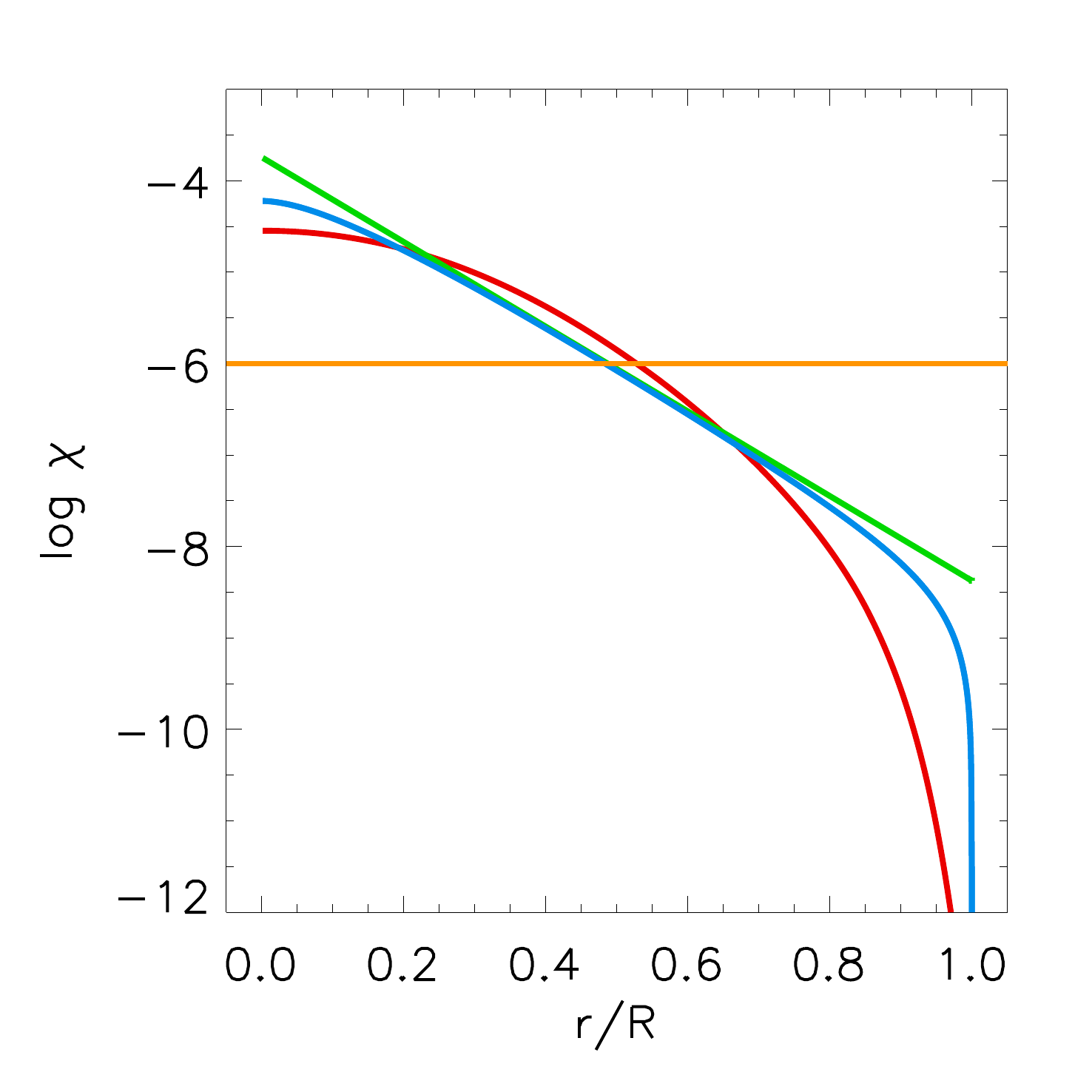}
  \includegraphics[scale=0.60]{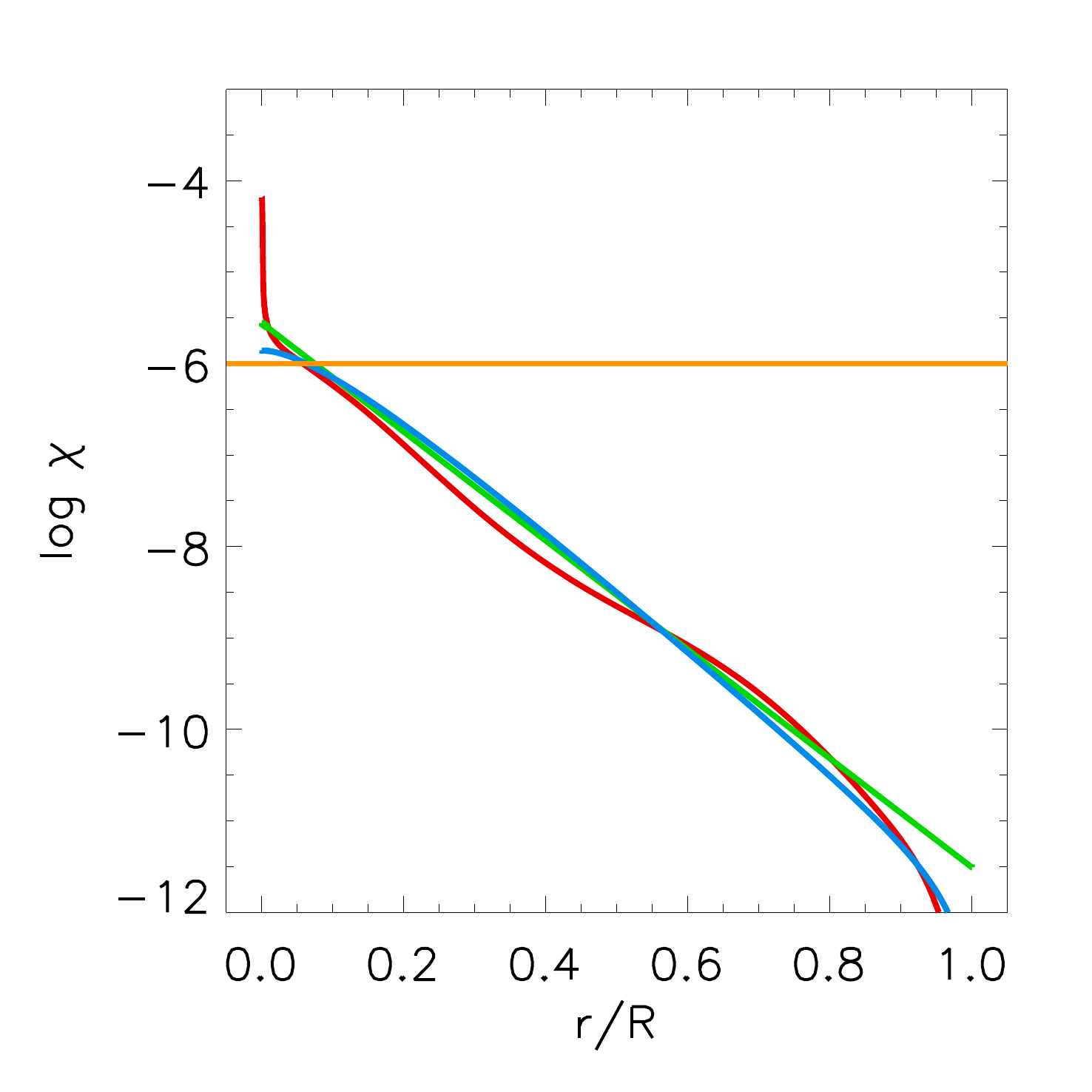}
  \caption{
Modified potential function function $\chi$ of eqn. (\ref{eqn:chipot}) (red) for the same MESA models as in Figure \ref{figure_3} and $15\; M_\odot$.  Left: ZAMS; right: BSG phase at \teff = 7900K. The analytic fits of eqn. (18) with $\zeta$ of (19) or (20)  are shown in blue and green, respectively. The screening parameter $\chi_c$ = 10$^{-6}$ is represented by the horizontal orange line. 
}
\label{figure_4}
\end{figure*}

Figure \ref{figure_3} displays an example for the density distribution fit of a 15 \Msun~star at the zero-age-mains-sequence (ZAMS). While the fit is not perfect, in particular above r/R $\sim$ 0.9, it is good enough for the discussion of the effects of modified gravity, as we will explain below. We use $\delta_{ms}$ = 0.08 and $f_{\delta}^{ms}$ = 2.5 for the main sequence density fits.

During the evolution away from the main sequence into the BSG phase the density distribution changes dramatically. The stellar core contracts and the outer layers expand.  As a result, the density increases strongly in a small central volume, while it decreases for the rest of the star. Figure \ref{figure_3} shows the example of 15 \Msun~star at the end of BSG phase at \teff = 7900K. A simple analytical fit is

\begin {equation}
  \rho(r) = \rho_c^{bsg},~~r \leq r_c, ~~r_c = f_{\delta}^{bsg}H, ~~H=\delta_{bsg} R,
\end{equation}
\begin{equation}
 \rho(r) = \rho_i^{bsg}e^{{r_c-r} \over H}~~~~~r_c \leq r \leq R,
\end{equation}

and

\begin{equation}
  \rho_i^{bsg} = {M - M_c \over {4\pi (\delta_{bsg} H)^3}}\times {1 \over \xi_{bsg}}
\end{equation}

with

\begin{equation}
  \xi_{bsg} = 2[(1+{r_c \over H} + {r_c^2 \over 2H^2}) - e^{r_c - R \over H}(1+{1 \over \delta_{bsg}} + {1 \over 2\delta_{bsg}^2})].
\end{equation}    

We apply $\delta_{bsg}$ = 0.06 as the best fit to describe the exponential decline outside the stellar core.
For the mass M$_c$ confined in the contracted central core we use the core mass of the  ZAMS phase defined as

\begin{equation}
  M_c = {4\pi \over 3} \rho_c^{ms} (f_{\delta}^{ms} \delta_{ms})^3 R_{ms}^3,
\end{equation}

where R$_{ms}$ is the stellar radius on the ZAMS. With R the actual stellar radius in the BSG phase we then adopt

\begin{equation}
  f_{\delta}^{bsg} = 0.2 f_{\delta}^{ms}{R_{ms} \over R} {\delta_{ms} \over \delta_{bsg}},
\end{equation}

which means that the core radius r$_c$ in the BSG phase is one fifth of the core radius on the ZAMS taking into account the contraction of the stellar core. This leads to the simple relationship between the central densities of the ZAMS and BSG phases

\begin{equation}
  \rho_c^{BSG} = 125 \rho_c^{ms}.
\end{equation}

\begin{figure}[ht!]
  \includegraphics[scale=0.6]{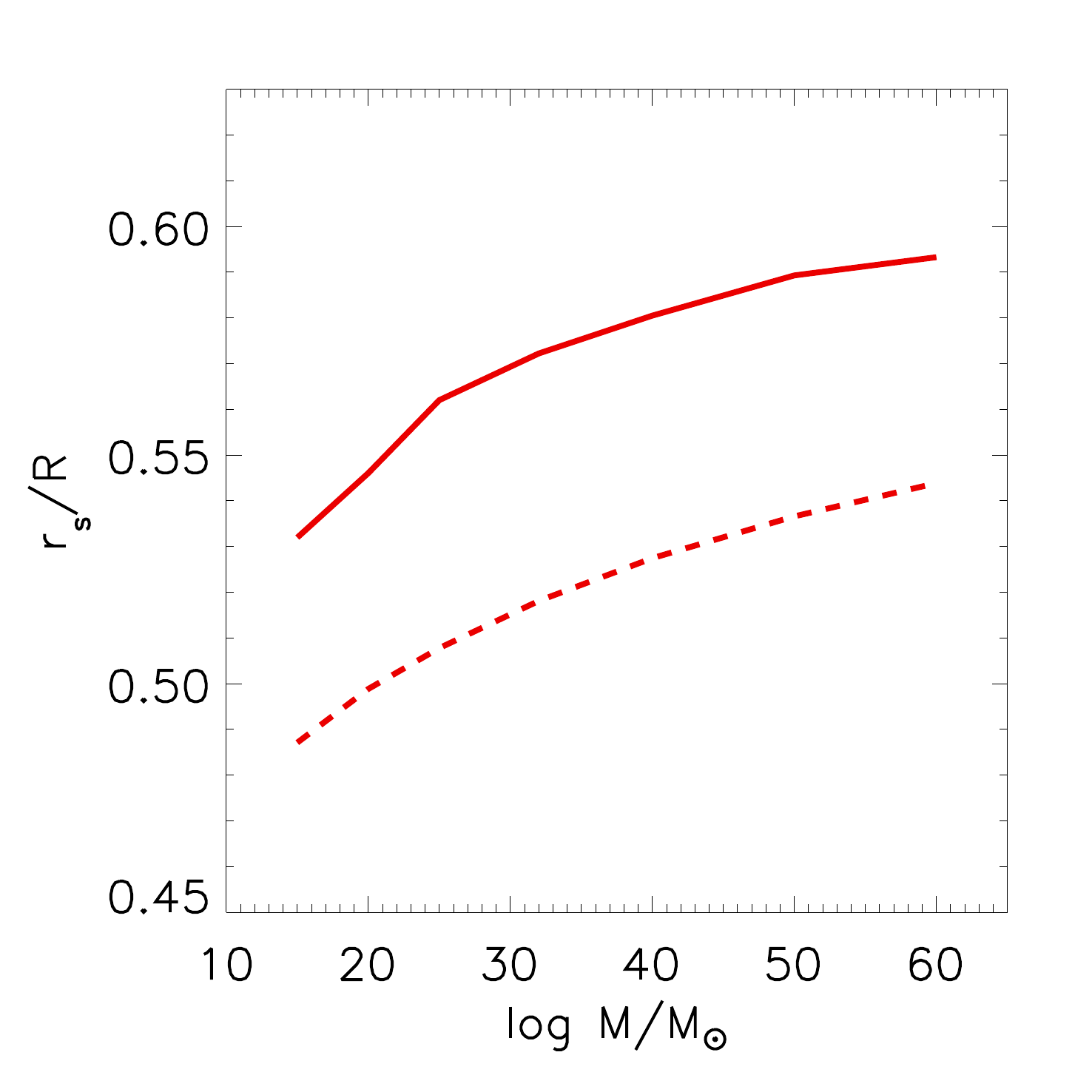}
  \caption{
Screening radius of MESA ZAMS models (red solid) as function of stellar mass for a screening parameter $\chi_c$ = 10$^{-6}$. The analytical approximation of eqn. (21) is shown as the dashed curve.     
}
\label{figure_5}
\end{figure}

\begin{figure}[ht!]
  \includegraphics[scale=0.6]{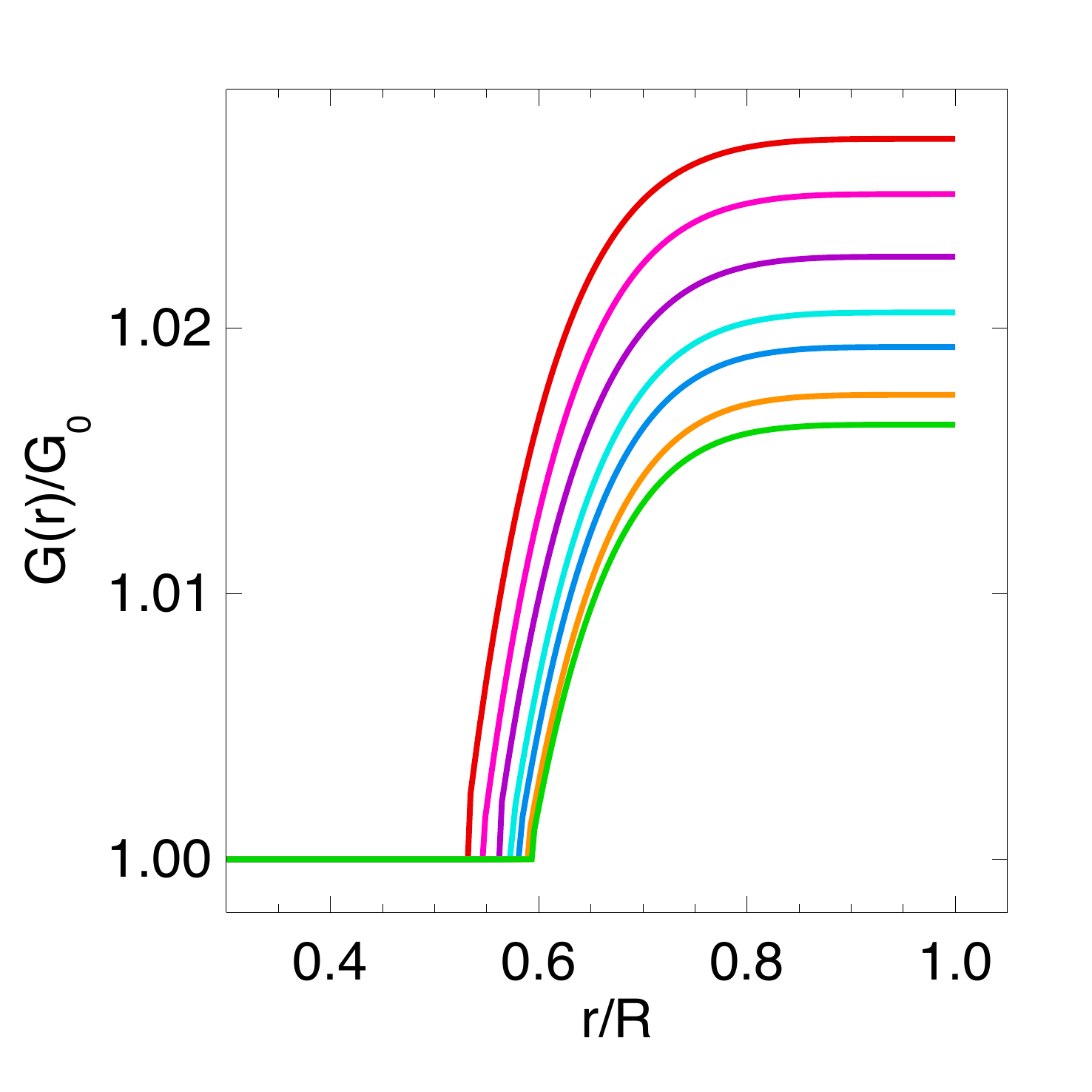}
  \caption{
Effects of modified gravity: G(r)/G$_0$ for MESA ZAMS  models with $\chi_c$ = 10$^{-6}$ and $\alpha_c$ = 1/3. Red: 15 \Msun, pink: 20 \Msun, violet: 25 \Msun, cyan: 32 \Msun, blue: 40 \Msun, orange: 50 \Msun, green: 60 \Msun.
}
\label{figure_6}
\end{figure}

The analytical fit obtained in this way is not excellent but to zero order captures the development of the density distribution during the evolution to the BSG phase well.

The analytical approximations of the density stratification inside the massive stars can now be used to calculate the screening radius using eqn. (4). We obtain

\begin{equation}
  \chi_c = {4\pi \over c^2}  G_0 \rho_0  \delta^2 R^2 \zeta({r_s \over \delta R})
\end{equation}

with

\begin{equation}
  \zeta({r_s \over \delta R}) = ({r_s \over \delta R} +1)e^{-{rs \over \delta R}} - ({1 \over \delta} + 1)e^{-{1 \over \delta}}.
\end{equation}.  

The density $\rho_0$ corresponds to $\rho_c^{ms}$e$^{r_c \over H}$ for the ZAMS and to $\rho_i$ in the BSG phase, respectively. For R and $\delta$ we select the corresponding ZAMS and BSG values.

In order to asses at which radius the interior of the the star is screened we study when the modified potential
\begin{equation}
\chi\left(\frac{r}{R}\right) \equiv \frac{4\pi}{c^2}G_0\int\limits_r^Rr\rho(r)\;dr    
\end{equation}
drops below the chameleon threshold for screening $\chi_c$.
Figure \ref{figure_4} shows the function $\chi$ of eqn. (20) versus r/R for the two examples of Figure \ref{figure_3} and compares with the analytical approximation. 
We see that for radii r/R $\leqq$ 0.8 the failure of the analytical density approximation in the main sequence case (left part of Figure \ref{figure_3}) has only a small influence on the analytical approximation of $\chi$(r). The reason is that the density in the outer stellar layers is small and, therefore, its contribution to the radial integral is of minor importance.

It is obvious from Figure \ref{figure_4} that even on the ZAMS a large fraction of the star would be unscreened against the effects of modified gravity if $\chi_c$ were as large as 10$^{-6}$. The effect becomes even more dramatic at the end of the BSG phase where all stellar layers except the very core are unscreened. This is the result of the core-halo density distribution inside BSGs as displayed in Figure \ref{figure_3} and described eqn. (11) and (12). We note that for  $\chi_c$ = 5$\times$10$^{-7}$ the horizontal orange line in Figure \ref{figure_4} would be 0.3 dex lower and the screening radii slightly larger.

In order to derive a simple analytical approximation for the screening radius we replace $\zeta$ of eqn. (19) by

\begin{equation}\label{eqn:chipot}
  \zeta_{app}({r_s \over \delta R}) = w \times 10^{-{r_s \over u\delta R}}
\end{equation}

with u = 2.7  and w = 3.0 for ZAMS stars and u = 2.8 and w = 2.0 for BSG stars. As can be seen from Figure \ref{figure_4} using $\zeta_{app}$ instead of $\zeta$ in eqn. (18) approximates the function $\chi$ equally well. Using $\zeta_{app}$ from eqn. (18) we can then approximate the screening radius by

\begin{equation}
  {r_s \over R} = u\delta \times \mathrm{log} ({w \over A})
\end{equation}

with

\begin{equation}
  A = {\chi_c \over {4\pi \over c^2}  G_0 \rho_0  \delta^2 R^2}.
  \end{equation}

\begin{figure}[ht!]
  \includegraphics[scale=0.6]{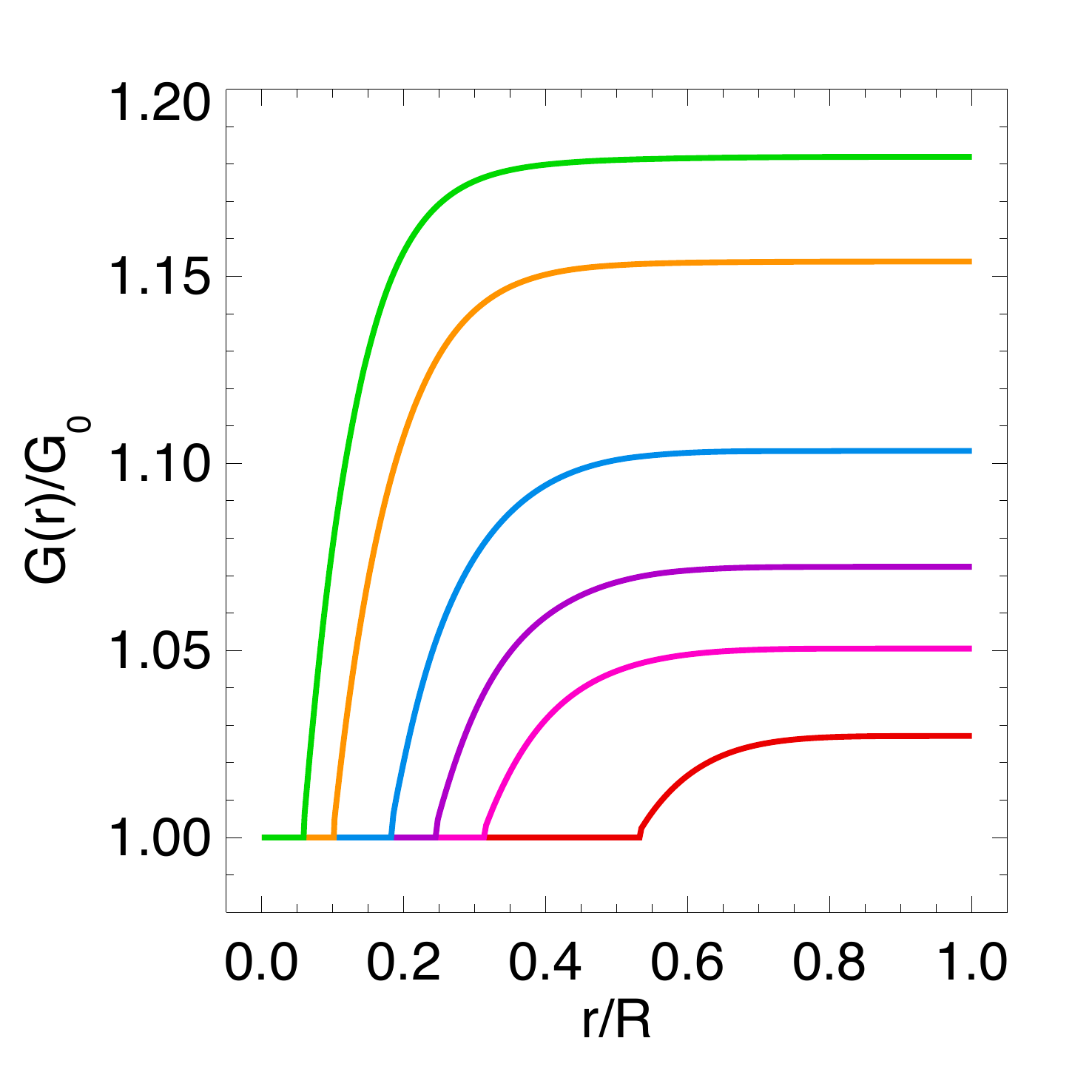}
  \caption{
Effects of modified gravity during the evolution of a 15 \Msun star: G(r)/G$_0$ for $\chi_c$ = 10$^{-6}$ and $\alpha_c$ = 1/3. Red: \teff = 32609K, pink: \teff = 24737K, violet: \teff = 19502 K, blue: \teff = 14684K, orange: \teff = 10046K, green: \teff = 7961K.
}
\label{figure_7}
\end{figure}
  
Figure  \ref{figure_5} shows the ZAMS screening radii r$_s$ obtained directly from numerical integral of eqn. (4) (corresponding to the red curve in Figure \ref{figure_3}) for different stellar masses compared with the approximation of eqn. (21). While there is a 9\% systematic offset, the approximation describes the trend with stellar mass very well. The screening radius moves further out when stellar mass increases. This is the result of ${1 \over A} \propto {M \over R}$ and the radius mass relationship of R $\propto$ M$^{0.57}$ on the ZAMS. This leads to ${1 \over A} \propto$ M$^{0.43}$ resulting in a larger value of r$_s$/R, when the ZAMS mass becomes larger.

  For BSG stars the ratio of  ${M \over R}$ is much smaller, because the stars have expanded significantly, and consequently the screening radii become as small as r$_s$/R = 0.06 at the end of the BSG phase as  indicated by Figure \ref{figure_4} for the example of 15\Msun. For evolutionary tracks with larger initial ZAMS masses r$_s$/R can reach values as small as 0.02.

\begin{figure*}[ht!]
  \includegraphics[scale=1.10]{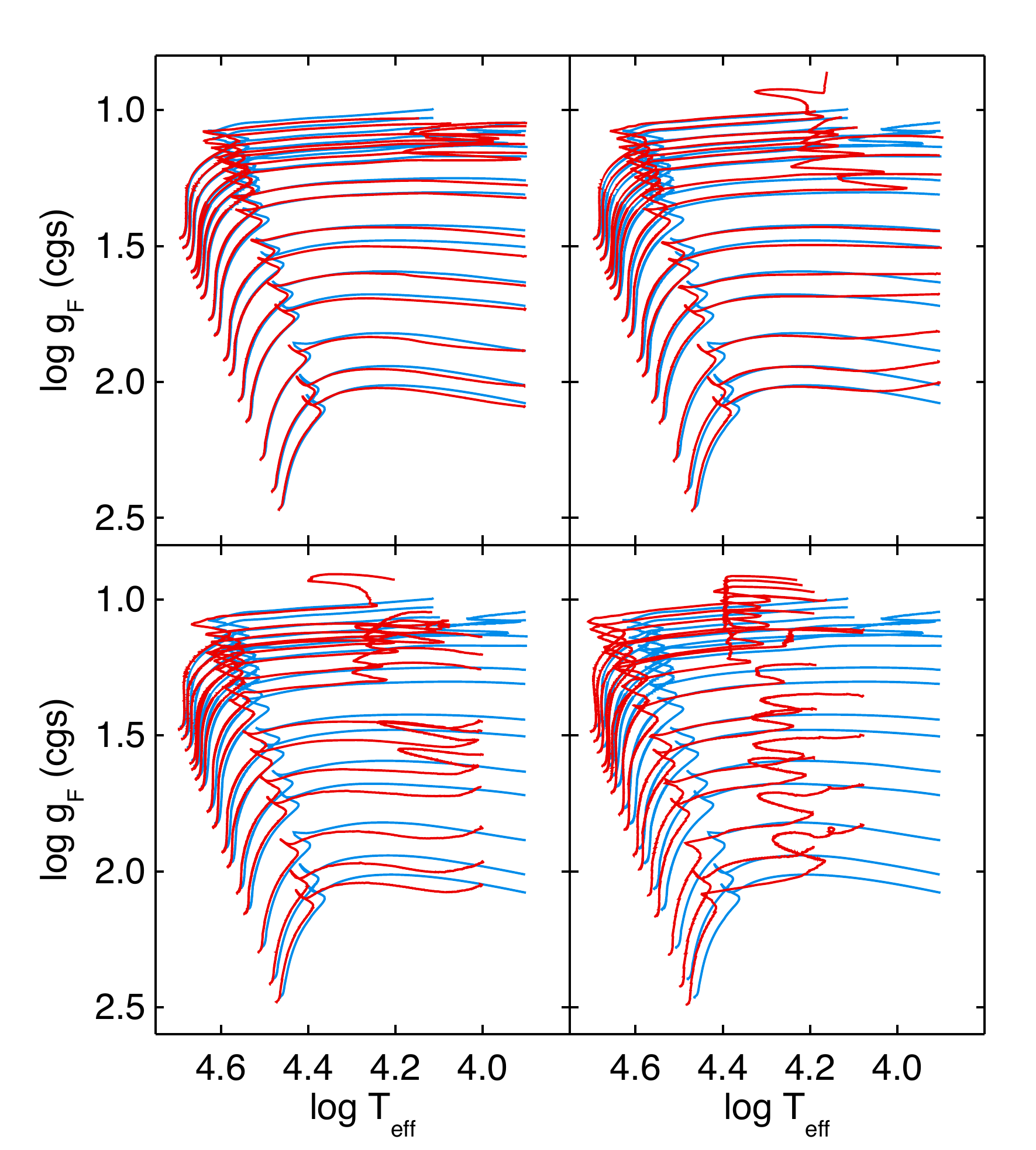}
\caption{
Stellar evolution in the (\teff, \loggf)-diagram. The same evolutionary tracks as in Figures \ref{figure_2} are shown. Top left: $\alpha_c$ = 1/3 and $\chi_c$ = 5$\times$10$^{-7}$, top right:
$\alpha_c$ = 1/3 and $\chi_c$ = 10$^{-6}$, bottom left: $\alpha_c$ = 1 and $\chi_c$ = 5$\times$10$^{-7}$, bottom right: $\alpha_c$ = 1 and $\chi_c$ = 10$^{-6}$.
}
\label{figure_8}
\end{figure*}
  
The location of the screening radius determines how strong the modification of the gravitational force is in the unscreened region inside the star through the ratio of M(r$_s$)/M(r)  in eqn. (5). Figure \ref{figure_6} shows the stratification of G(r)/G$_0$ for selected ZAMS models of different mass. We see that the effects of modified gravity increase with decreasing stellar mass, because the screening radius moves inward according to Figure \ref{figure_5} and eqn. (21) and, therefore, encircles a smaller mass M(r$_s$) relative to the total mass. In the same way modified gravity becomes more important in the evolution of a massive star from the ZAMS to the BSG phase, because the screening radius moves inward during the course of evolution. This is shown in Figure \ref{figure_7}  for the example of 15 \Msun.

The saturation of G(r)/G$_0$ towards increasing radii r/R  is a consequence of the low density in the outer stellar layers. This means that the incremental increase of the mass M(r) confined within radius r is very small when moving outward and is very close to the total stellar mass. With eq. (5) this leads to the saturation of  G(r)/G$_0$.

Stellar luminosity is affected by the modifications of gravity following eqn. (6). An increase of G(r) also enhances the luminosity and the effect becomes stronger when a larger fraction of the star is unscreened. This explains the qualitative behaviour in Figure \ref{figure_2}. On the main sequence changes in luminosity are larger at lower masses than at higher masses. They are very small for $\chi_c$= 5$\times 10^{-7}$ and $\alpha_c$ = 1/3 but increase  with $\chi_c$ and $\alpha_c$, because  the screening radius moves inward and G(r) increases. In the BSG phase the increases in luminosity follow the same trend but are significantly larger since a much smaller fraction of the star is unscreened. At $\chi_c$ = 10$^{-6}$ and also with $\alpha_c$ = 1 the effects become extreme and lead to a significant increase of the outer stellar envelopes, in particular, at lower masses. These dramatic changes of the outer stellar structure require extremely small time steps in the numerical calculation of the evolution to accomplish a converged solution. This is the reason why we have stopped the calculations at somewhat higher temperatures at values between 10000K to 15000K.

\begin{figure*}[ht!]
  \includegraphics[scale=0.60]{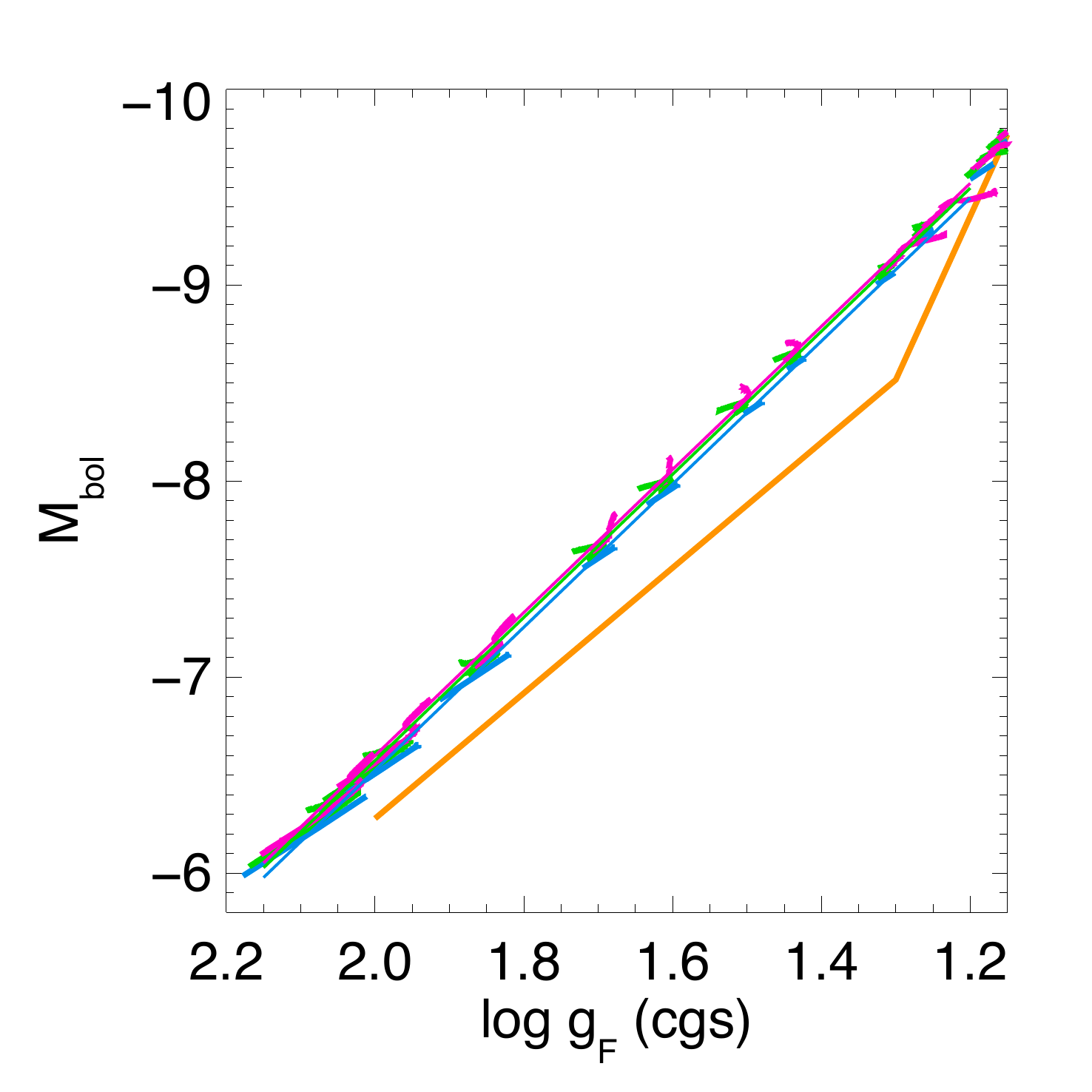}
  \includegraphics[scale=0.60]{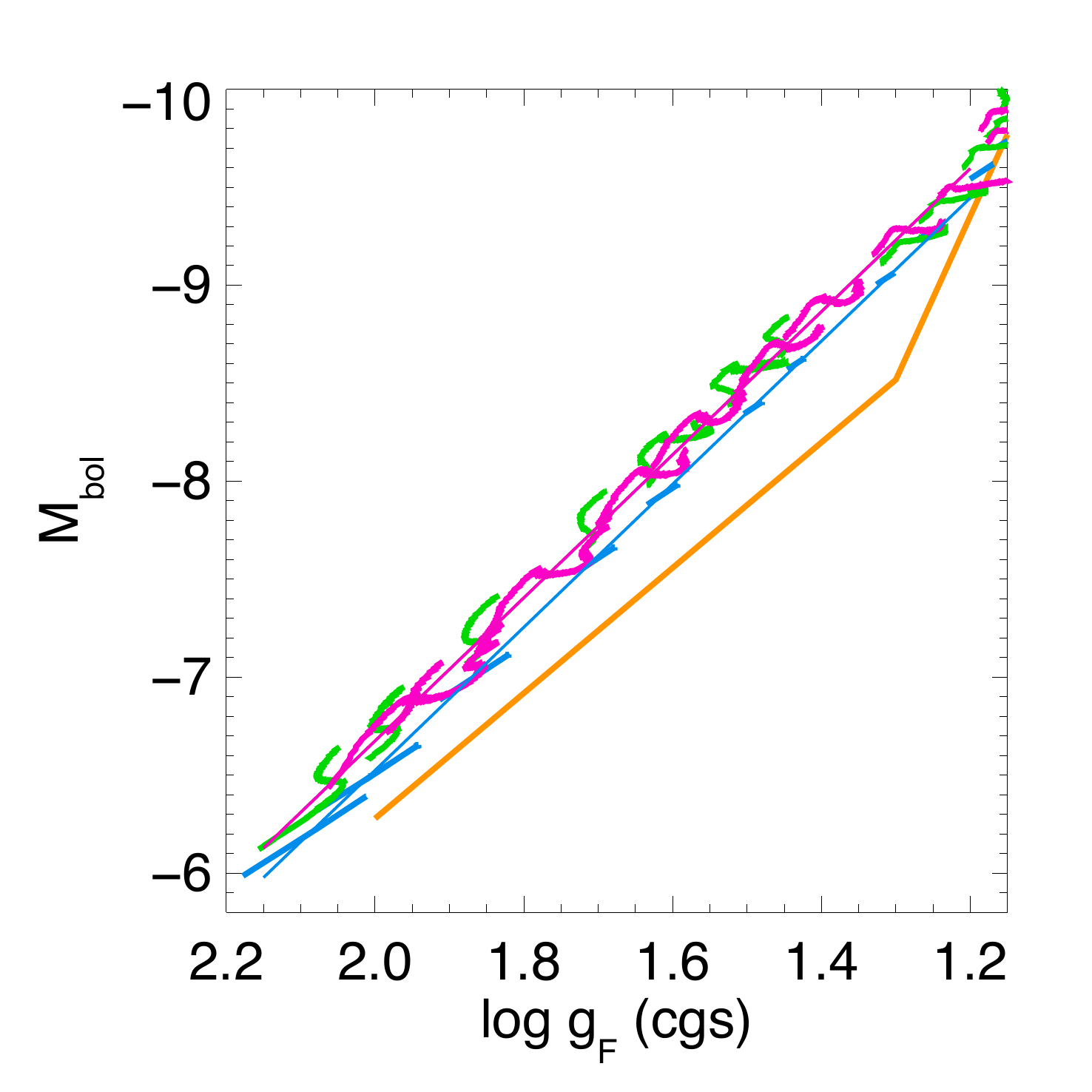}
\caption{
Stellar evolution in the (\Mbol, \loggf)-diagram. Evolutionary tracks in the \teff~range of BSG stars are shown for Newtonian gravity (blue) and modified gravity. Left: $\alpha_c$ = 1/3 and $\chi_c$ = 5$\times$10$^{-7}$ (green), $\chi_c$ = 10$^{-6}$ (pink). Right: $\alpha_c$ = 1 and $\chi_c$ = 5$\times$10$^{-7}$ (green), $\chi_c$ = 10$^{-6}$ (pink). The observed BSG FGLR (\citealt{Urbaneja2017}, see Figure \ref{figure_1}) is shown as an orange line. The straight blue, green and pink lines are simple FGLR fits to the stellar evolution calculations. For $\alpha_c$ =1 the fits for $\chi_c$ = 5$\times$10$^{-7}$ and 10$^{-6}$ coincide and only the fit for 10$^{-6}$ is visible in the plot (see text).
}
\label{figure_9}
\end{figure*}
  
\section{Modified gravity and flux-weighted gravity}

Figure \ref{figure_8} displays the spectroscopic Hertzsprung-Russell diagram (sHRD) as introduced by \cite{Langer2014}, where flux-weighted gravity is plotted versus effective temperature.  Again we compare stellar evolution models calculated with Newtonian and modified gravity. The evolutionary tracks of the Newtonian models demonstrate why flux-weighted gravity is a promising distance indicator. It stays roughly constant in the BSG evolutionary phase (where stars become very bright at visual wavelength because of the dependence of bolometric correction  on \teff)~ but at the same time it correlates strongly with luminosity.

The influence of modified gravity is two-fold. Since \gf~$\propto$ GM/L, an increase of G(r) through modified gravity increases \gf, whereas the simultaneous increase in luminosity may lead to a decrease of \gf. For $\chi_c$ = 5 $\times$10$^{-7}$ and $\alpha_c = 1/3$ the influence of G(r) is dominant, in particular for BSG at lower masses. This is still true, when $\alpha_c$ is increased to 1, at least at lower masses. For $\chi_c$ = 10$^{-6}$ the changes in luminosity start to dominate and \gf~ decreases relative to BSG models with Newtonian gravity. Generally, the effects of modified gravity are small at  $\chi_c$ = 5 $\times$10$^{-7}$ and $\alpha_c = 1/3$ but become much more pronounced when $\chi_c$ and $\alpha_c$ increase for the physical reasons already discussed in the previous section.

\begin{figure}[ht!]
  \includegraphics[scale=0.60]{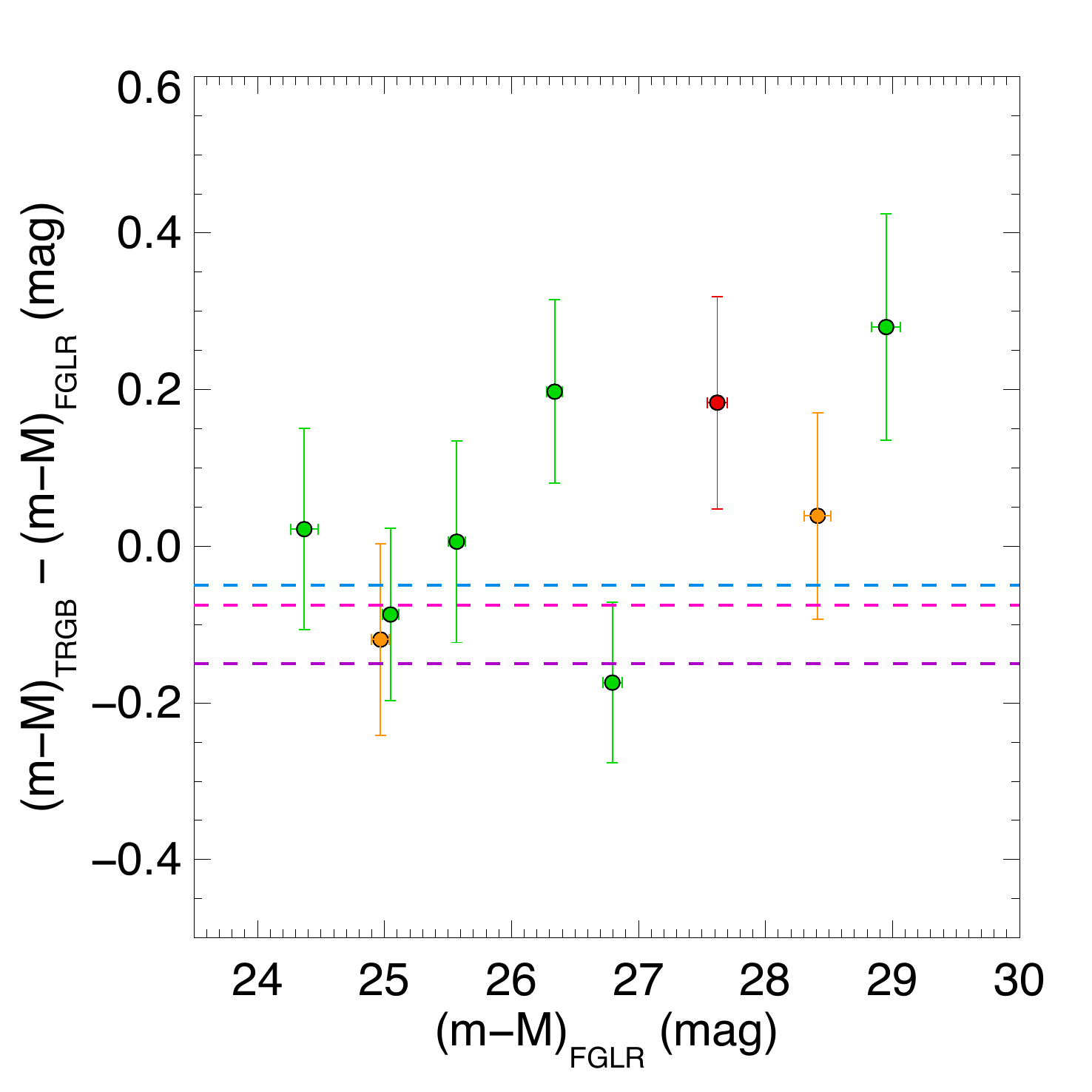}
\caption{
  Differences between observed TRGB and FGLR distance moduli as a function of FGLR distances modulus for 9 nearby galaxies. Galaxies unscreened against modified gravity for $\chi_c$ = 5 $\times$10$^{-7}$ and 10$^{-6}$ are plotted in green. Galaxies plotted in orange are shielded for $\chi_c$ = 5 $\times$10$^{-7}$ but unscreened for 10$^{-6}$. The galaxy shown in red is shielded at both values of $\chi_c$. \textbf{The predictions by modified gravity are shown as dashed horizontal lines (blue: $\alpha_c$ = 1/3, $\chi_c$ = 5 $\times$10$^{-7}$; pink: $\alpha_c$ = 1/3, $\chi_c$ = 10$^{-6}$; violet: $\alpha_c$ = 1, $\chi_c$ = 5 $\times$10$^{-7}$ and 10$^{-6}$).}
}
\label{figure_10}
\end{figure}

The effects of modified gravity encountered in Figure \ref{figure_8} must also be visible in the FGLR of the stellar evolution models. In Figure \ref{figure_9} we have transformed the evolutionary tracks of Figure \ref{figure_8} into diagrams, which displays bolometric magnitude against flux-weighted gravity. We also compare with the observations by \citet{Urbaneja2017} already shown in Figure \ref{figure_1}. Because of the comparison with BSG observations we restrict the plot of the tracks to the \teff~range of BSG.

Before we discuss the influence of modified gravity we need to comment on the comparison of the Newtonian tracks with the observations. At \textbf{\loggf~$\geq$ 1.3} we note an offset  in \Mbol. This offset was already noticed in the work by \citet{Meynet2015} (see their figures 3, 6, and 7), who compared stellar evolution models with observations based on the old FGLR calibration by \citet{Kudritzki2008}. \citet{Farrell2019} compared with the same set of observations but used models which included the effects of binary star evolution. They found a similar offset (see their figures 2 and 3). At this point it is unclear what the physical reason for the offset is. There are  many reasons which could affect the evolution of massive stars into the BSG phase such as changes in mass-loss or rotational properties (see \citealt{Meynet2015}) or changes in the internal angular momentum transport and rotational mixing (see \citealt{Taormina2020}). At the same time there is also the possibility that the spectroscopic measurement of BSG flux-weighted gravities are subject to systematic uncertainties.  We note that a systematic shift of 0.1 dex in \loggf~ at higher gravities of the observational data would resolve the problem. Such a shift is, of course, possible. For \loggf~$\leq$ 1.3 there is a strong disagreement between between the observed and stellar evolution FGLR slopes. So far, no explanation for this discrepancy has been found. We note, however, that this steep part of the FGLR is of little influence for extragalactic FGLR distance determinations, since most of the extragalactic BSG found are in the range of the higher gravity part. 

For a simple straightforward estimate of the potential influence of modified gravity on BSG distances determinations using the FGLR we ignore the discrepancy between observed and model FGLRs and concentrate on the discussion of the systematic differential effects on evolutionary tracks with modified gravity relative to the Newtonian tracks.

The differential effects for $\alpha_c$ = 1/3 in Figure \ref{figure_9} are small. We fit a straight FGLR curve to the evolutionary tracks assuming \Mbol = a(\loggf-1.5)+b with constant a=3.65 but different b for the three different sets of evolution models. We restrict the fit to the range 2.2 $\leq$ \loggf~$\leq$ 1.2. The difference $\Delta$b between modified gravity and the Newtonian models is -0.05 mag and -0.075 mag for $\alpha_c$ = 1/3 and $\chi_c$ = 5$\times$10$^{-7}$ and 10$^{-6}$, respectively. These differences are comparable to the distances modulus uncertainties obtained from a typical FGLR distance determination. But since the effects are systematic, they may become important, when larger samples of galaxies are used, for instance, to calibrate second order distance indicators. For  $\alpha_c$ = 1 the differences are even larger and amount to -0.15 mag for both values of $\chi_c$. An effect of this size would certainly be important for extragalactic distance determinations. As a consequence, FGLR distances would become larger with increased influence of modified gravity. The changes in distance are comparable to the ones found by \citet{Jain2013} for Cepheid stars.

So far, spectroscopic studies of BSG with FGLR measurements have been carried out for nine galaxies (see introduction and Table \ref{tab:table_1}). While this is a relatively small sample, it provides a first opportunity to investigate potential effects of a fifth force on this new stellar distance indicator. As shown by \citet{Jain2013} the fact that TRGB distances are not affected by modified gravity in the range of $\alpha_c$ and $\chi_c$ considered here can be used to observationally constrain these parameters. For this purpose it is important to distinguish between galaxies shielded or unscreened against modified gravity. As we already mentioned in the introduction the stars in many galaxies are shielded against modified gravity because of the galaxy gravitational potential or the superimposed potential of neighboring galaxies. However, isolated smaller galaxies are unscreened and here stars would be fully influenced by the fifth force of modified gravity. In those galaxies the FGLR shifts caused by modified gravity would change the distance determinations while TRGB distances would remain unchanged. 

\begin{deluxetable}{ccc}
\tabletypesize{\small}
\tablewidth{0pt}
\tablenum{1}
\tablecolumns{3}
\tablecaption{Galaxies with \textbf{observed} FGLR and TRGB distance moduli} \label{tab:table_1}
\tablehead{
 \colhead{name} & \colhead{(m-M)$_{\mathrm{FGLR}}$} & \colhead{(m-M)$_{\mathrm{TRGB}}$}\\
 \colhead{  } & \colhead{mag}    & \colhead{mag} }
\startdata
NGC3621 & 28.95$\pm0.11$ & 29.23$\pm$0.09 \\
M83         & 28.41$\pm0.11$ & 28.45$\pm$0.08 \\
M81         & 27.62$\pm0.08$ & 27.80$\pm$0.11 \\
NGC55     & 26.79$\pm0.08$ & 26.62$\pm$0.07 \\
NGC300   & 26.34$\pm0.06$ & 26.54$\pm$0.10 \\
NGC3109 & 25.57$\pm0.07$ & 25.58$\pm$0.11 \\
WLM         & 25.05$\pm0.06$ & 24.96$\pm$0.09 \\
M33         & 24.97$\pm0.07$ & 24.85$\pm$0.10 \\
IC1613     & 24.37$\pm0.11$ & 24.39$\pm$0.07 \\
\enddata
\vspace{-0.6cm}
\end{deluxetable}

The conditions for screening have been investigated by \citet{Cabre2012} and \citet{Jain2013}. Stars inside a galaxy are shielded against the fifth force, when the absolute value of the galaxy potential is larger than ${3 \over 2}\chi_c$. This is the case of internal screening. As shown by \citet{Jain2013} the observed maximal rotational velocity v$_m$ can be used to estimate the potential. This leads to the condition 2$\times$10$^{-7}$(v$_m$/200 kms$^{-1}$)$^2$ $\geq$ $\chi_c$ for internal screening. The condition for external screening can be approximated by the addition of the point source potentials of all neighbor galaxies within a radius $\lambda_c$ + r$_i$ through ${3 \over 2}{1 \over c^2}\sum {GM_i \over d} \geq \chi_c$. M$_i$, r$_i$ and d$_i$ are the dynamical masses, virial radii and distances of the neighbor galaxies. $\lambda_c$ is the Compton length of the fifth force and related to the screening parameter $\chi_c$ via $\lambda_c$ = 3200$\sqrt{\chi_c}$ Mpc. For the calculation of galaxy screening in the nearby universe \citet{Cabre2012} provide two galaxy catalogues on their website  based on the work by \citet{Karachentsev2004} and \citet{Lavaux2011}, which we have utilized. (We have updated some of the values of v$_m$ and M$_i$ taking into account more recent work).  

Applying the screening conditions as outlined we find that three galaxies of our sample (M81, M83, and M33) are screened against a fifth force characterized by $\chi_c$ = 5$\times$10$^{-7}$. The remaining galaxies (NGC3621, NGC300, NGC 55, NGC 3109, IC 1613, WLM) are unscreened. For $\chi_c$ = 10$^{-6}$ M33 and M83 become unscreened as well.

We have used the FGLR calibration of eqn. (1) and (2) to re-determine distances to the eight galaxies mentioned in the introduction. We also determined an FGLR distance to NGC 300 using the spectroscopic results by \citet{Kudritzki2008}. All nine galaxies also have accurate TRGB distances determined consistently in a homogeneous way and published in the EDD database (http://edd.ifa.hawaii.edu, see \citealt{Tully2009}). A subset of four galaxies (M81, NGC55, NGC300, NGC3109) has also TRGB distances determined by the ANGST project \citep{Dalcanton2009}. For those we use the mean between EDD and ANGST. The FGLR and TRGB distance moduli for the nine galaxies are given in Table \ref{tab:table_1}.

Figure \ref{figure_10} displays the differences $\Delta$ = (m-M)$_{\mathrm{TRGB}}$ - (m-M)$_{\mathrm{FGLR}}$  between TRGB and FGLR distance moduli for the shielded and unscreened galaxies in the sample. The mean value $\overline{\Delta}$ for all nine galaxies is $\overline{\Delta}$ = 0.039$\pm$0.052 mag, whereas for $\chi_c$ = 5$\times$10$^{-7}$   the mean for the unscreened galaxies is $\overline{\Delta}$ = 0.041$\pm$0.070 mag and $\overline{\Delta}$ = 0.035$\pm$0.087 mag for the shielded sample. For $\chi_c$ = 10$^{-6}$ we obtain  $\overline{\Delta}$ = 0.021$\pm$0.055 mag for the unscreened sample and $\Delta$ = 0.183$\pm$0.135 for the one remaining shielded galaxy M81.

According to the differential effects of our fits to the stellar evolution FGLRs of Figure \ref{figure_9}  we would  expect $\overline{\Delta}$ = -0.05 mag  and -0.75 mag for $\alpha_c$ = 1/3 and $\chi_c$ = 5$\times$10$^{-7}$ and 10$^{-6}$, respectively, and -0.15 mag for $\alpha_c$ = 1. However, the values obtained for the unscreened samples are positive and the differences are larger than 2.7 standard deviations for $\alpha_c$ = 1. Formally, adopting a Gaussian distribution for $\overline{\Delta}$ the probability of $\alpha_c$ = 1 being consistent with our measurement is smaller than 4\%. The probabilities for $\alpha_c$ = 1/3 and $\chi_c$ = 5$\times$10$^{-7}$ and 10$^{-6}$ are 11\% and 5\% , respectively.

We note that by calculating probabilities in this way we make the assumption that the distance moduli obtained by the TRGB and FGLR method, respectively, are not affected by systematic effects based on their calibrations. If, for instance,  $\Delta$ would have a systematic positive offset caused by calibration systematics, this would compensate for the negative shifts induced by modified gravity in the case of unscreened galaxies. Of course, by including the screened samples in our consideration we can study the differential effects between screened and unscreened galaxies. Unfortunately, the number of screened galaxies is small and the means are more uncertain.  For $\chi_c$ = 5$\times$10$^{-7}$ we obtain for $\widetilde{\Delta}$ = $\overline{\Delta}_{\mathrm{unscreened}}$ - $\overline{\Delta}_{\mathrm{screened}}$ = 0.006$\pm$0.111 mag. The value is still positive albeit with a large error. This means that in this statistically  more uncertain differential consideration $\alpha_c$ = 1/3 cannot be ruled out at this $\chi_c$ value. On the other hand, $\alpha_c$ = 1 can be ruled out to 92\%. For $\chi_c$ = 10$^{-6}$ a meaningful differential determination is not possible, because only the value for one galaxy is available for the screened sample.

Very obviously, increasing the number of shielded and unscreened galaxies with measured FGLR distances would make the result more significant. At this point, our results support the conclusions found by \citet{Jain2013} as described in section 3.


\section{Summary and conclusions}

We have constructed stellar evolution models for massive stars in the range from 12 \Msun~to 60 \Msun~accounting for the influence of modified gravity in the equation of hydrostatic equilibrium. We find an increase of stellar luminosity already on the main sequence and the effect becomes stronger in the BSG phase. The reason is the change in internal density stratification. BSG with their contracted core and strongly expanded envelope have a core-halo density structure, which leaves a large fraction of the stellar volume unscreened against a potential fifth force. In consequence, clear effects of modified gravity are encountered for the evolutionary tracks in the HRD and sHRD.

Using the evolutionary tracks in the BSG phase we can construct a theoretical FGLR diagram, where we display \Mbol~against \loggf. The comparison with observations in the LMC reveals an offset between the observed and theoretical FGLRs which was already noted in previous work and which may be caused by deficiencies of the spectral diagnostics leading to the observed FGLR or uncertainties of the stellar evolution treatment leading to BSG. However, the differential effects between the stellar evolution calculations with Newtonian gravity and modified gravity still allow for an estimate of the influence of modified gravity on FGLR distance determinations. Distance moduli would become 0.15 mag larger for a fifth force parameter $\alpha_c$ = 1 and shielding parameters $\chi_c$ = 5$\times$10$^{-7}$ and 10$^{-6}$. We use a comparison between observed TRGB distances, which should be unaffected by modified gravity, and FGLR distances in galaxies shielded and unscreened against the fifth force to constrain $\alpha_c$. If we assume that there is no systematic offset between TRGB and FGLR distances, then we find that $\alpha_c$ = 1 can be ruled out with 96\% confidence. For $\alpha_c$ = 1/3 modified gravity distance moduli would increase by 0.05 mag and 0.075 mag for $\chi_c$ = 5$\times$10$^{-7}$ and 10$^{-6}$, respectively. The constraints on modified gravity are slightly weaker in this case. $\chi_c$ = 5$\times$10$^{-7}$  is unlikely by 89\% and  $\chi_c$ = 10$^{-6}$ by 95\%. If we allow for a potential systematic offset between TRGB and FGLR distances, then the constraining results are more uncertain.  For $\chi_c$ = 5$\times$10$^{-7}$  $\alpha_c$ = 1 can still be ruled out to 92\% but $\alpha_c$ = 1/3 is unlikely only to 60\%. For $\chi_c$ = 10$^{-6}$ no constraints are possible, because the sample of shielded galaxies is too small. In summary, the results are comparable with the ones obtained by \citet{Jain2013} from a study of Cepheid stars, where  for $\alpha_c$ = 1/3 values of  $\chi_c \geqq$ 5$\times$10$^{-7}$ were be ruled out with 95\% confidence and for $\alpha_c$ = 1 values of $\chi_c \geqq$ 1$\times$10$^{-7}$ with similar evidence.

\acknowledgments
This work was initiated  and supported by the Munich Excellence Cluster Origins funded by the Deutsche Forschungsgemeinschaft (DFG, German Research Foundation) under Germany's Excellence Strategy EXC-2094 390783311. We thank Jeremy Sakstein for constructive engaged discussion and valuable input.\\

\bibliography{ms}

\end{document}